\documentclass[aps,twocolumn,superscriptaddress,nofootinbib,pra]{revtex4}
\usepackage{psfrag}
\usepackage{amssymb}
\usepackage{graphicx}

\def\be{\begin{equation}}
\def\ee{\end{equation}}
\def\bea{\begin{eqnarray}}
\def\eea{\end{eqnarray}}

\begin{document}

\title{Measuring the purity of a qubit state:
entanglement estimation with fully separable measurements}
\author{E.~Bagan}
\affiliation{Grup de F{\'\i}sica Te{\`o}rica \& IFAE, Facultat de Ci{\`e}ncies,
Edifici Cn, Universitat Aut{\`o}noma de Barcelona, 08193 Bellaterra
(Barcelona) Spain}

\author{M.~A.~Ballester}
\affiliation{Department of Mathematics, University of Utrecht, Box
80010, 3508 TA Utrecht, The Netherlands}

\author{R.~Mu{\~n}oz-Tapia}
\affiliation{Grup de F{\'\i}sica Te{\`o}rica \& IFAE, Facultat de Ci{\`e}ncies,
Edifici Cn, Universitat Aut{\`o}noma de Barcelona, 08193 Bellaterra
(Barcelona) Spain}

\author{O.~Romero-Isart}
\affiliation{Grup de F{\'\i}sica Te{\`o}rica \& IFAE, Facultat de Ci{\`e}ncies,
Edifici Cn, Universitat Aut{\`o}noma de Barcelona, 08193 Bellaterra
(Barcelona) Spain}
\date{\today}

\begin{abstract}
Given a finite number $N$ of copies of a qubit state we compute
the maximum fidelity that can be attained using joint-measurement
protocols for estimating its purity. We prove that in the
asymptotic $N\to\infty$ limit, separable-measurement protocols can
be as efficient as the optimal joint-measurement  one if classical
communication is used. This in turn shows that the optimal
estimation of the entanglement of a two-qubit state  can also be
achieved asymptotically with fully separable measurements. Thus, quantum memories 
provide no advantage in this situation. 
The relationship between our global Bayesian approach and the quantum
Cram\'er-Rao bound is also discussed.
\end{abstract}
 \maketitle

The ultimate goal of quantum state estimation is to determine the
value of the parameters that fully characterize a given unknown
quantum state. However, in practical applications,
a partial characterization 
is often all one needs. Thus, e.g., knowing the purity of a qubit
state or the degree of entanglement of a bipartite state may be
sufficient to determine whether it can perform some particular
task~\cite{white} ---See Ref.~\cite{gisin} for recent experimental progress on estimating the degree of polarization (the purity) of light beams. This paper concerns this type of situation.

To be more specific, assume we are given $N$ identical copies of
an unknown qubit mixed state $\rho(\vec r)$, so that the state of
the total system is $\rho^N(\vec r)\equiv[\rho(\vec r)]^{\otimes
N}$. The set of all such density matrices $\{\rho(\vec r)\}$ can
be mapped into the Bloch sphere ${\cal B}=\{\vec r :\ r\equiv|\vec
r|\le1\}$ through the relation $\rho(\vec r)=(\openone+\vec
r\cdot\vec\sigma)/2$, where
$\vec\sigma=(\sigma_x,\sigma_y,\sigma_z)$ is a vector made out of
the three standard Pauli matrices. Our aim is to estimate the purity, $r$, as
accurately as possible by performing suitable measurements on the
$N$ copies, i.e., on $\rho^N(\vec r)$. This problem can also be
viewed as the parameter estimation of a depolarizing
channel~\cite{depolarizing} when it is fed with $N$ identical states.

The estimation protocols are broadly divided into two classes
depending on the type of measurements they use: joint and
separable. The former treats the system of $N$ qubits as a whole,
allowing for the most general measurements, and leads to the most
accurate estimates or, equivalently, to the largest fidelity
(properly defined below). The latter, treats each copy separately
but classical communication can be used in the measurement
process. This class is particularly important because it is
feasible with nowadays technology and it offers an economy of
resources. In this paper we show that for a sufficiently large
$N$, separable measurement protocols for purity estimation can
attain the  optimal joint-measurement fidelity bound. The power of
separable measurement protocols in achieving optimal performance
has also been demonstrated in other
contexts~\cite{us-local,others, discrim}.

It has been shown~\cite{vidal} that given $N$ copies of a
bipartite qubit pure state, $|\Psi\rangle_{AB}$, the optimal
protocol for measuring its entanglement consists in estimating the
purity of $\rho(\vec r)\equiv{\rm
tr}_B(|\Psi\rangle_{AB}\langle\Psi|)$, where ${\rm tr}_B$ is the
partial trace over the Hilbert space of party $B$
(see~\cite{susana,horodecki} for related  work on bipartite mixed
states). We thus show that for {\em large $N$} this entanglement
can be optimally estimated by performing just {\em separable}
measurements on {\em one} party (party $A$ in this discussion) of
{\em each} of the $N$ copies of~$|\Psi\rangle_{AB}$.

Though many of our results here concern finite $N$, special
attention is paid to  the asymptotic regime, when $N$ is large.
There are several reasons for this. First, in this limit, formulas
greatly simplify and usually reveal important features of the
estimation protocol. Second, the asymptotic theory of quantum
statistical inference, which has become in recent years a very
active field in mathematical statistics~\cite{masahito-book},
deals with  problems such as the one at hand. Our results give
support to some quantum statistical methods for which only
heuristic proofs exist; e.g., the applicability  of the integrated
quantum Cram\'er-Rao bound in the Bayesian approach (which is
formulated below)~\cite{us-prep}.

In the first part of this paper we obtain the optimal joint
estimation protocols and the corresponding fidelity bounds. In
addition to the general case of states in $\cal B$, which was
partially addressed in~\cite{vidal}, we also discuss the situation
when the unknown state is constrained  to lie on the equatorial
plane $\cal E$ of the Bloch sphere $\cal B$. In the second part,
we discuss separable measurement protocols, we prove that they
saturate the joint-measurement bound asymptotically and we state our
conclusions.

Mathematically, the problem of estimating the purity of $\rho(\vec
r)$ can be formulated within the Bayesian framework as follows (see~\cite{keyl} for an alternative
approach).
Let ${\cal R}_{\cal O}=\{R_\chi\}$ be the set of estimates of $r$,
each of them  based on a particular outcome $\chi$ of some
generalized measurement, $\cal O$, over $\rho^N(\vec r)$. In full
generality, we assume that such measurement is  characterized by a
Positive Operator Valued Measure (POVM), namely, by a set of
positive operators ${\cal O}=\{O_\chi\}$ that satisfy $\sum_\chi
O_\chi=\openone$ ($\chi$ can be a continuous variable, in which
case the sum becomes an integral over $\chi$).
A separable measurement is a particularly interesting instance of a POVM for
which each $O_\chi$ is a tensor product of $N$ individual operators
(usually projectors) each one of them acting on $\rho(\vec r)$.
Next, a figure of merit, $f(r,R_\chi)$, is introduced as a
quantitative way of expressing the quality of the purity
estimation. Throughout this paper we use
\begin{eqnarray}
f(r,R_\chi)&\equiv&2\max_{\vec m} \left[{\rm tr}\sqrt{\rho^{1/2}(\vec r)\rho(R_\chi \vec m)
\rho^{1/2}(\vec r)}\right]^2-1\nonumber\\
&=&rR_\chi+\sqrt{1-r^2}\sqrt{1-R_\chi^2}={\bf r}\cdot{\bf R}_\chi,
\label{fidelity}
\end{eqnarray}
%
%
where $|\vec m|=1$, i.e., $[1+f(r,R_\chi)]/2$ is the standard
fidelity~\cite{fuchs} (see also \cite{fid}) between $\rho(\vec r)$
and $\rho(R_\chi \vec n)$, where we have defined $\vec n=\vec
r/r$. 
Throughout this paper we refer to $f(r,R_\chi)$ also as
fidelity for short. Its values are in the range $[0,1]$, where unity corresponds to perfect determination. It is interesting to note that in Uhlmann's geometric representation of the set of density matrices as the hemisphere $(1/2){\mathbb S}^3\subset{\mathbb R}^4$, the function $D(r,R_\chi)=(1/2)\arccos f(r,R_\chi)$ is the geodesic (Bures) distance~\cite{som}
between two sets  (two parallel 2-dimensional spheres) characterized by the purities~$r$ and~$R_\chi$ respectively.

In the same spirit as
in~\cite{us-prep,alberto}, we have written $f(r,R_\chi)$ as a
scalar product of the two unit vectors ${\bf a}=(\sqrt{1-a^2},a)$;
$a=r, \,R_\chi$. The optimal protocol is obtained by maximizing
\begin{equation}
F({\cal O},{\cal R}_{\cal O})=\sum_\chi\int d\rho f(r,R_\chi) {\rm
tr}[\rho^N(\vec r) O_\chi], \label{averaged fidelity}
\end{equation}
where $d\rho$ is the prior probability distribution of $\rho(\vec
r)$, and we identify the trace as the probability of obtaining the
outcome $\chi$ given that the state we measure upon is
$\rho^N(\vec r)$. Thus, $F$ is the average fidelity. The
maximization is over the estimator (guessed purity) ${\cal
R}_{\cal O}$ and the POVM ${\cal O}$. Using Schwarz inequality the
optimal estimator is easily seen to be
\begin{equation}
R_\chi^{\rm opt}={V_\chi\over\sqrt{{\bf V}_\chi\cdot{\bf
V}_\chi}}; \quad {\bf V}_\chi=\int d\rho\; {\bf r} \, {\rm
tr}[\rho^N(\vec r) O_\chi], \label{optimal guess}
\end{equation}
and
\begin{equation}
F({\cal O})\equiv \max_{\{{\cal R}_{\cal O}\}}F({\cal O},{\cal
R}_{\cal O})=\sum_\chi \sqrt{{\bf V}_\chi\cdot{\bf V}_\chi} \  .
\label{optimal fidelity}
\end{equation}
We are still left with the task of computing $F^{\rm
max}=\max_{{\cal O}} F({\cal O})$.

In this formulation, we need to provide a prior probability
distribution (prior for short) $d\rho$, which encodes our initial
knowledge about $\rho(\vec r)$. Here we assume to be completely
ignorant of both $\vec n$ and $r$. Our lack of knowledge about the
former is properly  represented with the choice $d\rho \propto
d\Omega$ (solid angle element), which states that {\em \`a priori}
$\vec n$ is isotropically distributed on ${\cal B}$. Therefore, we
write
\begin{equation}
d\rho={d\Omega\over4\pi} w(r)dr;\quad \int_0^1dr\, w(r)=1.
\label{measure}
\end{equation}
While there is  wide agreement on this respect, the $r$-dependence
of the prior is controversial and so far  we will not stick to any
particular choice. 
Nevertheless, it is worth keeping in mind that  the hard sphere prior $w(r)=3 r^2$ shows 
up in the context of entanglement estimation~\cite{zycz},
whereas the Bures
prior $w(r)=(4/\pi) r^2 (1-r^2)^{-1/2}$ is most
natural in connection with distinguishability of density matrices~\cite{fuchs,fid,prior}.

We are now in a position to compute $F^{\max}$. We first assume no
constraint on $\cal O$, thus allowing for the most general
measurement setup. The density matrix $\rho^N(\vec r)$ can be
written in a block-diagonal form, where each block,
$\rho_{Nj\alpha}(\vec r)$, transforms with a corresponding
spin~$\bf j$ irreducible representation of $SU(2)$ and $\alpha$
($\alpha=1,2,\dots, n_j$) labels the different $n_j$ occurrences
of the same block~\cite{cirac, us-prep}. This implies that  each
element, $O_\chi$, of the optimal POVM can be likewise chosen to
have the same block-diagonal structure. 
%

Given a POVM $\tilde{\cal O}$ of this type, we consider the two-stage  measurement protocol ${\cal O}$ consisting of
({\em i}\hspace{.1em})~a `preliminary'  measurement  of
the projection of the state $\rho^N(\vec r)$ onto the $SU(2)$
irreducible subspaces, followed by ({\em ii\hspace{.1em}})~the measurement defined by~$\tilde{\cal O}$. 
The outcomes of $\cal O$ are thus labeled by three indexes $\chi=(j,\alpha,\xi)$, and the corresponding operators are defined by $O_{j\alpha\xi}=\openone_{j\alpha}{\tilde O}_\xi\openone_{j\alpha}$.
Since the
projector on each irreducible subspace, $\openone_{j\alpha}\equiv\sum_m |jm;\alpha\rangle
\langle jm;\alpha|$,
commutes with $\rho^N(\vec r)$, the probabilities ${\rm
tr}[\rho^N(\vec r)\, \tilde O_\xi]$ are the marginals of ${\rm
tr}[\rho^N(\vec r)\, O_{j\alpha\xi }]$ and the fidelity cannot decrease by using $\cal O$ instead of the original~$\tilde{\cal O}$. 
In our quest for optimality, we thus stick to these two-stage measurements.

We next recall that $\rho(\vec r)=U\rho(r \vec z)U^\dagger$ for a
suitable $SU(2)$ transformation $U$, where $\vec z$  is the unit
vector along the $z$ axis, and that $d\Omega$ can be replaced by
the Haar measure of $SU(2)$. Using Schur's lemma the integral
in~(\ref{optimal guess}) gives
\begin{equation}
{\bf V}_{j\alpha\xi}={{\rm tr}(
O_{j\alpha\xi})\over2j+1}\int dr\,w(r)\,{\bf r}\,{\rm tr}
[\rho_{Nj\alpha}(r\vec z)] . \label{Vjchi}
\end{equation}
Hence, the estimate $R^{\rm opt}_{\chi}=R^{\rm opt}_{j\alpha\xi}$ turns out to be
independent of the outcomes~$\xi$ (of~$\tilde{\cal O}$), and we can write $R^{\rm
opt}_{j\alpha}$ instead. This, in turn, renders the maximization
in~(\ref{optimal fidelity}) trivial, since, using the relation $\sum_\xi
O_{j\alpha\xi}=\openone_{j\alpha}$, we see that the right
hand side of~(\ref{optimal fidelity}) becomes also independent of
$\tilde {\cal O}$, and we can drop the
subscript $\xi$ from now on.

The bottom line is that, assuming an isotropic prior, the optimal
purity estimation is entirely based on the outcomes of $\cal I$
(no additional information about the purity can be extracted from the
state) and we might as well choose not to perform any further
measurement ($\{\tilde O_\xi\}\to\openone$). With this choice, the
prefactor in~(\ref{Vjchi}) becomes unity.  Since the $n_j$ spin
$\bf j$ blocks $\rho_{Nj\alpha}$ all give an identical
contribution
\begin{equation}
{\rm tr}[\rho_{Nj\alpha}(r\vec z)]=\sum_{m=-j}^j p_r^{{N\over2}-m}
q_r^{{N\over2}+m}, \label{trace}
\end{equation}
where $p_r=(1-r)/2$, $q_r=1-p_r$,  the left hand side of~(\ref{Vjchi})
can be simply called ${\bf V}_{j}$,

 The maximal fidelity is
thus given by
\begin{equation}
F^{\rm
max}=\pmatrix{N\cr{N\over2}-j}{2j+1\over{N\over2}+j+1}\sum_j\sqrt{{\bf
V}_j\cdot{\bf V}_j} \ , \label{Fmax}
\end{equation}
where the coefficient in front of the sum
is~$n_j$~\cite{cirac,us-prep}. This, along with~(\ref{trace})
and~(\ref{Vjchi}), provides an explicit expression of~$F^{\rm
max}$. For large $N$, this can be computed~to~be~\cite{details}
\begin{equation}
F^{\rm max}= 1-{1\over2N}+ o(N^{-1}) . \label{Fasymp}
\end{equation}
One can also check that at leading order in $1/N$ the optimal
guess is $R^{\rm opt}_j=2j/N$, as one would intuitively expect.
These asymptotic results hold for any prior $w(r)$.

\begin{figure}
 \psfrag{N}{$N$}
 \psfrag{P}[bc]{$N(1-F^{\rm max})$}
 \includegraphics[width=8cm]{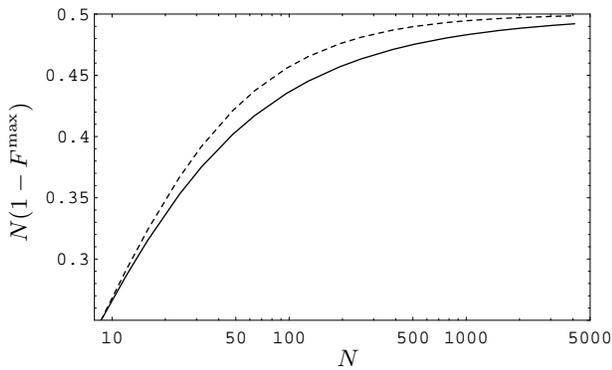}
 \caption {A log-linear plot of $N(1-F^{\rm max})$ in terms of the number  $N$ of copies  for
 the optimal joint measurement  and for the Bures (solid line)
 and hard sphere (dashed line) priors.}\label{fig}
 \end{figure}

In Fig.~\ref{fig}, we plot $N(1-F^{\rm max})$ as a function of $N$
in the range $10$--$5000$ for states in $\cal B$ and for  the Bures (solid line)
and the
hard sphere (dashed
line) priors. The two lines are seen to approach the
asymptotic value $1/2$ [which can be read off from
Eq.~(\ref{Fasymp}) ] for large~$N$ at a similar rate.

It is also interesting to analyze the case where $\vec r$ is known
to lie on the equatorial plane $\cal E$. With this information,
the prior probability distribution becomes
$d\rho=(d\phi/2\pi)w(r)dr$, where $\phi$ is the polar angle of the
spherical coordinates. Though it is still possible to use the
block-diagonal decomposition discussed above, the individual
blocks are now reducible under the unitary symmetry
transformations on~$\cal E$, i.e., under a $U(1)$ subgroup of
$SU(2)$. 
In full analogy to the general case,
the optimal POVM is given by the set
of one-dimensional projectors over the $U(1)$-invariant subspaces, $\{\openone_{j\alpha
m}\equiv|jm;\alpha\rangle\langle jm;\alpha|\}$,
and, as above, the equivalent representations,
labelled by $\alpha$,  contribute a multiplicative factor~$n_j$.
The analogous of~(\ref{trace}) is now
\begin{equation}
[\rho_{Nj\alpha}(r\vec x)]_{mm}=\!\!\sum_{m'=-j}^j \left[{\rm
d}_{mm'}^{(j)}(\mbox{${\pi\over2}$})\right]^2 p_r^{{N\over2}-m'}
q_r^{{N\over2}+m'}, \label{trace2D}
\end{equation}
where ${\rm
d}_{mm'}^{(j)}(\mbox{${\beta}$})$ are the standard Wigner d-matrices~\cite{edmonds}.
From~(\ref{trace2D}) we can compute ${\bf V}_{jm}$ and $F^{\rm max}$, as
in~(\ref{Fmax}), where in this case the sum extends over $j$ and
$m$. 
The resulting
expression can be evaluated for small $N$ but it is not very
enlightening. The corresponding plots for the
analogous of Bures and hard sphere priors are indistinguishable
from those in~Fig.~\ref{fig}. Far more interesting is the large $N$ regime. It
turns out that $F^{\rm max}$ is also given by~(\ref{Fasymp}) and
the optimal guess becomes $m$ independent, $R^{\rm
opt}_{jm}=2j/N+\dots$. Therefore, we see that the information
about~$\vec n$ becomes irrelevant in the asymptotic limit.
%
%

A word regarding quantum statistical inference is in order here.
It is often argued that the quantum Cram\'er-Rao
bound~\cite{holevo} can be integrated to provide an attainable
asymptotic lower bound for some averaged figures of merit, such as
the fidelity~(\ref{fidelity}).  Ours is a so-called one parameter
problem for which the quantum Cram\'er-Rao bound takes the simple
form ${\rm Var}\, R\ge H^{-1}(\vec r)/N$, where  ${\rm Var}\,
R\equiv\langle (R_\chi-\langle R_\chi\rangle)^2\rangle$ is the
variance of the estimator $R_\chi$, the average is over the
outcomes $\chi$ of a measurement,  $H(\vec r)$ is the quantum
information matrix~\cite{holevo}, and $R_\chi$ is assumed to be
unbiased: $ \langle R_\chi\rangle=r $. In our case $H(\vec
r)=(1-r^2)^{-1}$, and the bound is attainable. This provides in
turn an attainable asymptotic upper bound for the
fidelity~(\ref{fidelity}), since $\langle
f(r,R_\chi)\rangle\approx
1-\raisebox{.12em}{\mbox{\tiny$1\over2$}}H(\vec r)\,{\rm Var}\,
R+\dots$. Assuming one can integrate  this relations over the
whole of~$\cal B$ (including the region $r\approx1$, where $H(\vec
r)$ is singular), with a weight function given by the
prior~(\ref{measure}), we obtain Eq.~(\ref{Fasymp}).
Unfortunately, there are only heuristic arguments supporting this
assumption, but so far no rigorous proof exists in the
literature~\cite{van-trees}.


We now abandon the joint protocols to dwell on separable
measurement strategies for the rest of the paper.  Here we focus
on the asymptotic regime, but some brief comments concerning small
$N$ can be found in the conclusions.

In previous work~\cite{alberto}, some of the authors showed that
the maximum fidelity one can achieve in estimating both $r$ and
$\vec n$ (full estimation of a qubit mixed state) assuming the
Bures prior and using tomography
behaves as
\begin{equation}
F^{\rm max}_{\rm full}=1-{\xi\over N^{3/4}}+o(N^{-3/4})   ,
\label{Ffull}
\end{equation}
where $\xi$ is a positive constant. The same behavior one should
expect for our fidelity $F^{\rm max}$, since the effect of the
purity estimation is dominant in~(\ref{Ffull}). This strange power
law, somehow unexpected on statistical grounds, is caused by the
behavior of $w(r)$  in a small region \mbox{$r\approx 1$}. Indeed,
it is not difficult to convince oneself that if $w(r) \propto
(1-r^2)^{-\lambda}\approx 2(1-r)^{-\lambda}$ for $r\approx 1$, one
should expect $1-F^{\rm max}\propto N^{\lambda/2-1}+\dots$, for
$0<\lambda<1$ (for $\lambda=0$, hard sphere prior, one should
expect  logarithmic corrections). This differs drastically
from~(\ref{Fasymp}) which, as stated above, holds for
{\rm any} such values of $\lambda$. Would classical communication
be enough to restore the right power law $N^{-1}$ for $1-F^{\rm
max}$ and, moreover, saturate the bound of the optimal joint protocol?

On quantum statistical grounds, one should expect a positive
answer to this question since  the quantum Cram\'er-Rao bound is
attained by a separable protocol consisting in performing the (von
Neumann) measurements ${\cal M}=\{(\openone\pm\vec
n\cdot\sigma)/2\}$ on each copy. Note, however, that $\cal M$
depends on $\vec n=\vec r/r$, which is, of course, unknown {\em
\`a priori}. This protocol can only make sense if we are ready to
spend a fraction of the $N$ copies of $\rho(\vec r)$ to obtain an
estimate of~$\vec n$, use this classical information to design
$\cal M$ and, finally, perform this adapted measurement on the
remaining copies. This protocol was successfully applied to pure
states by Gill and Massar in~\cite{gill-massar}. We extend it to
purity estimation below.

Let us consider a family of priors of the form
\begin{equation}
w(r)={4\over\sqrt\pi}{\Gamma(5/2-\lambda)\over\Gamma(1-\lambda)}{r^2
(1-r^2)^{-\lambda}}  , \label{gen prior}
\end{equation}
which includes both the Bures ($\lambda=1/2$) and the hard sphere ($\lambda=0$) metrics. Despite
of this particular $r$ dependence, the final results apply to any
prior whose behavior near $r=1$ is given by~(\ref{gen prior}).

We now proceed {\em \`a la} Gill-Massar~\cite{gill-massar} and
consider the following one-step adaptive protocol: we take a
fraction $N^\alpha\equiv N_0$ ($0<\alpha<1$) of the $N$ copies
of $\rho(\vec r)$ and we use them to estimate $\vec n$. Tomography
along the three orthogonal axis $x$, $y$ and $z$, together with a
very elementary estimation based on the relative frequencies of
the outcomes~\cite{us-local},  enables us to estimate $\vec n$
with an accuracy given by
\begin{equation}
{\langle\Theta^2_r\rangle\over
2}\approx1-\langle\cos\Theta_r\rangle={3\over N_0}\left({1\over
r^2}-{1\over5}\right)+o(N_0^{-1}), \label{Theta}
\end{equation}
where $\Theta_r$ is the angle between $\vec n$ and its estimate.
Here and below $\langle\cdots\rangle$ is not only the average over
the outcomes of this tomography measurements, but also contains an
integration over the prior angular distribution $d\Omega/(4\pi)$
for fixed $r$. We see from~(\ref{Theta}) that the pure state limit is
\mbox{$\langle\Theta_{r\to1}^2\rangle\approx24/(5N_0)+\dots$}, and
one can compute the fidelity, as defined in~\cite{us-local}, to
check that it agrees with the result therein. This concludes the
first step of the protocol.

In a second step, we measure the projection of $\vec\sigma$ along
the estimated $\vec n$ obtained in the previous step. We perform
this von Neumann measurement on each of the remaining $N-N_0\equiv
N_1$ copies of the state $\rho(\vec r)$. We estimate the purity to
be $R=2N_+/N_1-1$,  where $N_\pm/N_1$ is the relative frequency of
$\pm1$ outcomes, and we drop the $N_+$ dependence of $R$ to
simplify the notation.

Obviously, as a random variable and for large $N_1$, $R$~is normally
distributed as $R\sim{\rm N}(r c_r,\sqrt{1-r^2
c^2_r}/\sqrt{N_1})$, where $c_r=\cos\Theta_r$. Hence, for large
$N_0$ and $N_1$ it makes sense to expand $f(r,R)$,
Eq.~(\ref{fidelity}), around $R= r c_r$,  and thereafter, because
of~(\ref{Theta}), expand the resulting expression around~$c_r=1$.
We obtain
\begin{equation}
F(r)= 1-{1\over 2
N_1}+{r^2\over1-r^2}\left({\langle\Theta^2_r\rangle\over4N_1}-{\langle\Theta^4_r\rangle\over8}\right)+\dots
, \label{<f>}
\end{equation}
where $F(r)$ is the average fidelity for fixed $r$, i.e., $\int
dr\,w(r) F(r)=F$. In view of~(\ref{Theta}),
$\langle\Theta_r^4\rangle\sim N_0^{-2}=N^{-2\alpha}$. Hence, the
two terms in parenthesis in~(\ref{<f>}) can only be dropped if
$\alpha>1/2$. Provided $w(r)$ vanishes as in~(\ref{gen prior})
with $\lambda<0$, we can integrate $r$ in~(\ref{<f>})  over the
unit interval to obtain
\begin{equation}
F=1-{1\over2N(1-N^{\alpha-1})}+o(N^{-1}) , \label{F in I}
\end{equation}
and we conclude that this protocol attains asymptotically the
joint-measurement bound~(\ref{Fasymp}).

However, most of the physically interesting priors~\cite{prior,zycz}, $w(r)$, not
only do not vanish as $r\to1$, but often diverge like~(\ref{gen
prior})  with $0<\lambda<1$. In this case (\ref{<f>}) cannot be
integrated, as the last term does not lead to a convergent
integral. This signals that the series expansion around $c_r=1$
leading to~(\ref{<f>}) is not legitimated in the whole of~$\cal
B$.

To fix the problem, we split $\cal B$ in two regions. A sphere
of radius $1-\epsilon$, $\epsilon>0$, which we call ${\cal B}^{\rm
I}$, and a spherical sheet  of thickness $\epsilon$:  ${\cal
B}^{\rm II}=\{\vec r: 1-\epsilon<r\le 1\}$. The fidelity can thus
be written as the sum of the corresponding two contributions:
$F=F^{\rm I}+F^{\rm II}$. While $F^{\rm I}$ can be obtained by
simply integrating~(\ref{<f>}) over ${\cal B}^{\rm I}$, where this
expansion is valid, some care must be taken in the region~${\cal B}^{\rm II}$. There, we proceed
as follows.

We compute the fidelity as if all the states in ${\cal B}^{\rm II}$ had the lowest possible purity ($r=1-\epsilon$)
when the first-step tomography was performed. This leads to a lower bound for $F^{\rm II}$,
because the lower the purity of a state  the less accurately  $\vec n$ can be determined [see Eq.~(\ref{Theta})], and hence, the worse its purity can be estimated in the
second step. 
The trick, which amounts to replacing $c_r$ by $c_{1-\epsilon}$, enables us to perform the $r$-integration prior
to $\langle\cdots\rangle$. We simply expand $f(r,R)$,
Eq.~(\ref{fidelity}), around $R= r c_{1-\epsilon}$ to obtain
\begin{eqnarray}
F(r)&\gtrsim& \Bigg\langle\sqrt{(1-r^2)(1-r^2c^2_{1-\epsilon})}\nonumber\\
&-&{1\over2N_1}\sqrt{{1-r^2\over 1-r^2 c_{1-\epsilon}^2}} +\dots\Bigg\rangle
,
\end{eqnarray}
where the dots stand for additional terms that are irrelevant to
the problem we are addressing here. Integrating this expression
and expanding around $c_{1-\epsilon}=1$ we obtain
\begin{eqnarray}
&&\kern-4em\int_{1-\epsilon}^1\kern-1.3em dr\,w(r) F(r)\gtrsim 1-{1\over2N_1}-
k_\lambda\left\langle(1-c_{1-\epsilon})^{2-\lambda}\right\rangle\nonumber\\
&&\kern3.4em-\left(1-{1\over2N_1}\right)\int_0^{1-\epsilon}\kern-1.3em
dr\,w(r)+\dots,
\end{eqnarray}
where $
k_\lambda={2^{2-\lambda}\Gamma({5\over2}-\lambda)\Gamma({3\over2}-\lambda)\Gamma(\lambda-2)/[\pi\Gamma(1-\lambda)]}
$. Putting together the different pieces of the calculation we
have
\begin{equation}
F\gtrsim 1-{1\over2N_1}-2^{\lambda-2}k_\lambda
\langle\Theta_{1-\epsilon}^2\rangle^{2-\lambda} +\dots,
\label{fidelity ok}
\end{equation}
$0<\lambda<1$, where now we can safely take the limit
$\epsilon\to0$. We see that by choosing 
\begin{equation}
{\rm
\max}\left\{{1\over2},{1\over2-\lambda}\right\}<\alpha< 1
\label{alpha}
\end{equation}
 we ensure that the joint-measurement
bound~(\ref{Fasymp}) is attained. It is worth
emphasizing that the last term in~(\ref{fidelity ok}), which is
completely missing in~(\ref{F in I}), is actually the dominant
contribution if $\alpha<1/(2-\lambda)$. For $\lambda=0$ we have
\begin{equation}
F^{\rm hard}\gtrsim1-{1\over2N_1}-{3   \langle\Theta^2_1\rangle
\log\langle\Theta^2_1\rangle\over8N_1}+\dots  ,
\end{equation}
and we again conclude that the protocol presented here attains the
joint-measurement bound. 

Two comments about the choice of $\alpha$ are in order.  First, numerical simulations show that the optimal value of $\alpha$ is very close to the lower bound in~(\ref{alpha}). Second, we see that the lower bound in~(\ref{alpha}) 
increases with increasing~$\lambda$. This can be understood by recalling that for large $N$,  the estimated purity $R$
is normally distributed with a variance of ${\rm Var}\,R=(1-r^2 c_r^2)/N_1$. For $\lambda\ll1$, the prior is a rather flat  function of $r$ and, on average, ${\rm Var}\,R=a/N_1$, where $a$ is a constant. Increasing
the accuracy by which $\vec n$ is determined  does not improve significantly the estimation of~$r$. Hence, using a small fraction of the number of copies  at the first stage of the protocol should be enough. 
This suggest that $\alpha$ must be relatively small.  In contrast, for $\lambda\approx1$ the prior peaks at $r=1$ and ${\rm Var}\,R=\Theta^2_r/N_1$. Hence, it pays to
spend a large fraction of $N$ to estimate $\vec n$ with high accuracy (as this drastically reduces ${\rm Var}\,R$), for which we need that $\alpha\approx 1$.

At this point one may wonder if the conclusions above depend upon
our particular choice of figure of merit. To get a grasp on this,
it is worth using again the standard pointwise approach to quantum
statistics. There, one is interested in the mean square error
${\rm MSE}\,R=\langle(R-r)^2\rangle$ for fixed $r$, where now the
average $\langle\cdots\rangle$ is over the outcomes of {\em all}
measurements for a fixed $\vec r$. One can write ${\rm
MSE}\,R={\rm Var}\, R + (\langle R\rangle-r)^2 $, where the second
term is the \emph{bias}. Using the same one-step adaptive protocol
described above, we get that the mean square error after step two
is
\begin{equation}
{\rm
MSE}\,R=\frac{H^{-1}(r)}{N_1}+{r^2\over4}\langle\Theta_r^4\rangle+
\dots .
\end{equation}
As above, the last term can be dropped if $\alpha>1/2$,
and
\begin{equation}
{\rm MSE}\,R=\frac{H(r)^{-1}}{N} +o[N^{-1}],
\end{equation}
saturating the quantum Cram\'er-Rao bound. This protocol is,
therefore, also asymptotically optimal in the present context.
Though the argumentation above is somehow heuristic,  it can be
made fully rigorous~\cite{ballester}.


In summary, we have addressed the problem of optimally estimating
the purity of a qubit state of which $N$ identical copies are
available. The optimal estimation of the entanglement of a
bipartite qubit state can be reduced to this problem. Though the
absolute bounds for the average fidelity involve joint
measurements, these bounds can be obtained asymptotically with
separable measurements. This requires classical communication
among the sequential von Neumann measurements performed on each of
the $N$ individual copies of the state. This result, which has
been speculated on quantum statistical grounds, is here proved for
the first time by a direct calculation. This leads to a very
surprising result: in the asymptotic limit of many copies,
bipartite entanglement, a genuinely non-local property, can be
optimally estimated by performing fully separable measurements.
This meaning that measurements can be performed not only on copies
of {\em one} of the two entangled parties, but on {\em each} of
these copies {\em separately}. This avoids the necessity of quantum memories.

For finite (but otherwise arbitrary) $N$, finding the optimal
separable measurement protocol is an open problem. Interestingly
enough, a `greedy' protocol designed to be optimal at each 
measurement step~\cite{us-local, others} leads to an unacceptably
poor estimation.  Notice that in the one-step adaptive  protocol
described above, part of the copies were spent (`wasted'  from a
`greedy'  point of view) in estimating $\vec n$.
We have seen that this strategy pays in the long run.
However, the `greedy' strategy optimizes measurements in the short run,
which translates into measuring $\vec\sigma$ along the same arbitrarily fixed axis on each copy of~$\rho(\vec r)$.
This yields a low value for the fidelity, which does not
even converge to unity in the strict limit $N\to\infty$.
This counterintuitive behavior of  the `greedy' protocol also appears in
other contexts as, e.g., economics, biology or social sciences
(see~\cite{parrondo} for a nice example).

We acknowledge useful conversations with Antonio Ac\'{\i}n, Richard Gill  and Juanma
Parrondo. This work is supported by the  Spanish Ministry of
Science and Technology project BFM2002-02588, CIRIT project
SGR-00185, Netherlands Organization for Scientific Research NWO,
the European Community projects QUPRODIS  contract no.
IST-2001-38877 and RESQ contract no IST-2001-37559.

\newcommand{\PRL}[3]{Phys.~Rev. Lett.~\textbf{#1}, #2~(#3)}
\newcommand{\PRA}[3]{Phys.~Rev. A~\textbf{#1}, #2~(#3)}
\newcommand{\JPA}[3]{J.~Phys. A~\textbf{#1}, #2~(#3)}
\newcommand{\PLA}[3]{Phys.~Lett. A~\textbf{#1}, #2~(#3)}
\newcommand{\JOB}[3]{J.~Opt. B~\textbf{#1}, #2~(#3)}
\newcommand{\JMP}[3]{J.~Math.~Phys.~\textbf{#1}, #2~(#3)}
\newcommand{\JMO}[3]{J.~Mod.~Opt.~\textbf{#1}, #2~(#3)}


\begin{thebibliography}{99}
\bibitem{details} The techniques required to compute Eq.~(\ref{Fmax}) are explained in detail in~\cite{us-prep}.
\end{thebibliography}
\end{document}